\documentclass[sigconf]{acmart}
\let\labelindent\relax
\usepackage{enumitem}

\setlist[enumerate]{wide=\parindent}

\usepackage{lipsum}

\usepackage{lmodern}
\usepackage{amssymb,amsmath}
\usepackage{ifxetex,ifluatex}
\usepackage{fixltx2e} 
\ifnum 0\ifxetex 1\fi\ifluatex 1\fi=0 
  \usepackage[T1]{fontenc}
  \usepackage[utf8]{inputenc}
\else 
  \defaultfontfeatures{Ligatures=TeX,Scale=MatchLowercase}
\fi
\IfFileExists{upquote.sty}{\usepackage{upquote}}{}
\IfFileExists{microtype.sty}{%
\usepackage{microtype}
\UseMicrotypeSet[protrusion]{basicmath} 
}{}
\hypersetup{
            pdftitle={MXNET-MPI: Embedding MPI parallelism in DL DAGs for scaling Deep Learning},
            pdfauthor={; Amith R Mamidala},
            pdfborder={0 0 0},
            breaklinks=true}
\usepackage{url}
\usepackage{listings}
\usepackage{graphicx,grffile}
\makeatletter
\def\maxwidth{\ifdim\Gin@nat@width>\linewidth\linewidth\else\Gin@nat@width\fi}
\def\maxheight{\ifdim\Gin@nat@height>\textheight\textheight\else\Gin@nat@height\fi}
\makeatother
\setkeys{Gin}{width=\maxwidth,height=\maxheight,keepaspectratio}
\IfFileExists{parskip.sty}{%
\usepackage{parskip}
}{
\setlength{\parindent}{0pt}
\setlength{\parskip}{6pt plus 2pt minus 1pt}
}
\ifx\paragraph\undefined\else
\let\oldparagraph\paragraph
\renewcommand{\paragraph}[1]{\oldparagraph{#1}\mbox{}}
\fi
\ifx\subparagraph\undefined\else
\let\oldsubparagraph\subparagraph
\renewcommand{\subparagraph}[1]{\oldsubparagraph{#1}\mbox{}}
\fi

\let\origthelstnumber\thelstnumber
\makeatletter
\newcommand*\Suppressnumber{%
  \lst@AddToHook{OnNewLine}{%
    \let\thelstnumber\relax%
    \advance\c@lstnumber-\@ne\relax%
  }%
}
\newcommand*\Reactivatenumber{%
  \lst@AddToHook{OnNewLine}{%
    \let\thelstnumber\origthelstnumber%
    \advance\c@lstnumber\@ne\relax}%
}
\begin{document}

\title{Efficient Embedding of MPI Collectives in MXNET DAGs for scaling Deep Learning}
 \date{}
 \author{Amith R Mamidala}
\affiliation{IBM T J Watson Research Center\\
 Yorktown Heights, New York, USA}
\email{amithr@us.ibm.com}



\begin{abstract}

Availability of high performance computing infrastructures such as clusters of GPUs and CPUs have fueled the growth of distributed learning systems. Deep Learning frameworks express neural nets as DAGs and execute these DAGs on computation resources such as GPUs. In this paper, we propose efficient designs of embedding MPI collective operations into data parallel DAGs. Incorrect designs can easily lead to deadlocks or program crashes. In particular, we demonstrate three designs: Funneled, Concurrent communication and Dependency chaining of using MPI collectives with DAGs. These designs automatically enable overlap of computation with communication by allowing for concurrent execution with the other tasks. We directly implement these designs into the KVStore API of the MXNET. This allows us to directly leverage the rest of the infrastructure. Using ImageNet and CIFAR data sets, we show the potential of our designs. In particular, our designs scale to 256 GPUs with as low as 50 seconds of epoch times for ImageNet 1K datasets.

\end{abstract}
\maketitle

\section{Introduction}\label{sec:introduction}
 As Deep Learning(DL) continues its dominance in a multitude of disciplines such as Image Classification, Speech Recognition and Natural Language Processing, the need for innovation in DL systems of scale to reduce training times gains utmost importance. 
In almost all of the DL frameworks, the neural network is expressed as a computation DAG with nodes representing the operators of the model and the edges indicating the flow of data. 
The DNN training in a worker comprises of the feed forward phase where the operators are exercised using the existing model weights and a back propagation phase where these  weights are updated using the gradients computed in this phase. Stochastic Gradient Descent (SGD) is the popular algorithm used to update the weights from the computed gradients. Frameworks such as MXNET, TensorFlow use symbolic programming to do the training. This approach is well suited for DAG executions. In particular, the computation is not immediately carried out but rather the DAG execution is optimized for efficiency in memory usage and runtime. Further, all the tasks of the DAG can be asynchronously executed by multiple independent execution units such as GPUs or CPUs providing efficient utilization of all the resources.  

For data parallel SGD, a Parameter Server(PS) is typically used
where the gradients are forwarded from workers to one or more servers which aggregate them. Though the PS is a suitable model for cloud computing as it is inherently fault tolerant, it doesn't utilize all the communication links of the network. In particular, future generation machines such as Sierra and Summit\cite{coral} would deploy thousands of nodes featuring  multiple NVIDIA GPUs per node interconnected by fast InfiniBand networks. MPI has been proven to deliver scalable performance on these supercomputers owing to the rich subset of collective algorithms designed to aggressively utilize all the available network bandwidth. Parallel DL training can take advantage of these collective algorithms by appropriately embedding them in the DAG executions of the workers.

It is non-trivial to efficiently incorporate global collective operations in the DAGs. Unlike imperative programs the execution of tasks in symbolic DAGs is triggered by the completion of data flow dependencies. This inherent asynchronicity allows the operation for a given gradient to be issued in a different order from another worker. Naively substituting the local operation with a global collective operator can easily lead to deadlocks or incorrect execution flows.  The PS design, in particular MXNET, overcomes this issue by "naming" each gradient and the servers aggregating all the gradients independently by the same "name". However, this cannot be translated directly to MPI calls as these operate on buffers within \textbf{communicators} and are agnostic to how these buffers are actually "named".  Further, according to MPI semantics operations for the same communicator have to be issued in order from the same communicator, posing a design challenge. Moreover, these operations can be overlapped with the other independent tasks in the back propagation phase. We explore all these issues and propose multiple solutions to the problem.  Importantly, our MPI Collective + DAG design patterns explained in the paper ensure optimal training performance, by automatically permitting overlap of collective communication progress with the rest of the DAG execution and at the same time overcoming these  constraints.



Specifically, our contributions in this paper are the following:
\begin{itemize}[leftmargin=*]
\item{We describe different designs for embedding the global MPI aggregation operator,  MPI\_Allreduce, in the symbolic execution of the DNN DAGs. Though we use MXNET, the problem of embedding is generic and the solutions can be adapted to any DAG based frameworks such as TensorFlow.}

\item{We propose three design alternatives: a) MPI Funneled ("Funnel")  b) MPI Dependency chaining ("DepCha") and c) MPI concurrent communicators ("ConCum"). In "Funnel" the main thread does all the communication while the other tasks are executed by separate threads of the DAG execution engine. In contrast,"DepCha" offloads a chain of collective operations to the engine that are ordered by artificially inserting dependencies across them. In "ConCom", multiple MPI communicators are used to concurrently execute the collective communications.}

\item{ We implement these designs using MXNET framework  and study the benefits of these approaches on CIFAR10 and ImageNet 1K~\cite{imagenet} data sets. We show the benefits of scaling as well as the improvements obtained due to adding concurrency and overlap in the collective operations.}

\item { Using large scale ImageNet 1K~\cite{imagenet} we demonstrate the key benefits of our approaches, namely better scaling by reducing the epoch times to as low as 50 seconds using over 256 MPI processes using one GPU per process and achieving more than 0.71 validation accuracies for the entire data set.}
\end{itemize}

\begin{figure*}[h]
       \includegraphics[width=\textwidth]{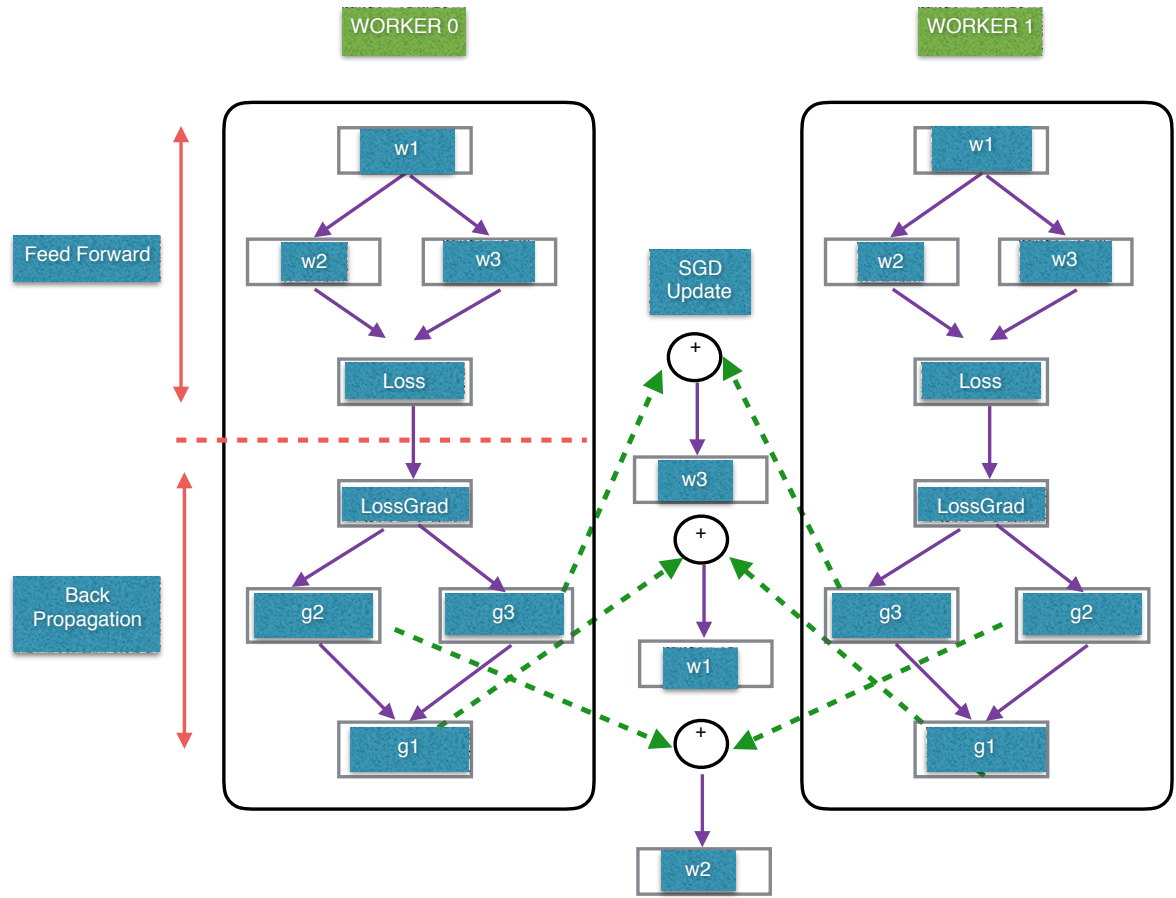}

    \vspace{-1mm}
  \caption{Parallel Dag Execution}
  \label{fig:parallel-dag}
\end{figure*}%
\section{Background \& Motivation}
\label{sec:motivation}

In this section, we first provide a brief overview of mini-batch parallel SGD which is the core algorithmic component of the training. We then describe the challenges in using global MPI aggregation operator, with parallel DAGs. This motivates the need for the different design explorations in this paper.
\subsection{Parallel SGD}\label{sec:parallel-sgd}
In a mini-batch SGD, the entire data is divided into several mini-batches, collectively known as the ``epoch". The computation iterates over the epoch, one mini-batch at a time. The model parameters at iteration t, $w_{t}$ are updated by an increment $\Delta w$ to get the parameters for the next iteration.
\begin{equation}\label{eq:param-update}
w_{t + 1} = w_{t} + \Delta w
\end{equation}
$\Delta w$ is computed as $\Delta w  = - \eta g$,
where $g$ is the gradient, $\eta$ is a hyper parameter called
as the learning rate. For the deep learning models, the model parameters and gradients are associated with the different layers of the network. The gradients are obtained after doing a forward pass and then an auto-differentiation in the backward step. The final gradient, $g$ is the average of all the gradients obtained from the data samples in a mini-batch. Also, since the gradients are obtained as soon as a backward step for a layer is computed, the model update can be done in parallel with the remaining backward phase computation. 

In parallel synchronous SGD, the mini-batch is divided across all the workers who wait for the global aggregation  of locally computed gradients. This value is multiplied by a scaling factor=1/$(mini\_batch\_size)$ and used for computing the next set of parameters, $w_{t+1}$, equation~\ref{eq:param-update}. 

\subsection{MPI Aggregation Operator}
Due to the exponential rise of compute flops, communication quickly becomes the bottleneck in the global aggregation. Modern interconnection networks such as InfiniBand offer very high bandwidth paths across the nodes. Unlike PS approaches, MPI collectives such as MPI\_Allreduce offer excellent performance.

This global operator does distributed aggregation across all the processes in the same group or more commonly known as MPI \textbf{communicator}. There is a default communicator, MPI\_COMM\_WORLD created every time MPI is initialized. For almost all of the applications, this single communicator is adequate to meet all the communication demands. An important requirement of using the single communicator is to issue the operations in the same order across all the MPI processes. This is because the internal collective algorithms implicitly "tag" all the messages using this order and issuing them in any other manner will cause deadlocks or incorrect behavior.
As we see below, this imposes a serious constraint in using them in parallel DAG aggregations.

\subsection{Naive MPI usage in parallel DAGs}
Figure~\ref{fig:parallel-dag} illustrates two DAGs constructed off a simple neural net comprising of three layers, run on two GPUs. The second layer comprises of two operators that can be run in parallel. Such constructions are common in image classification network architectures such as GoogleNet, inception-BN. As shown in the figure, the DAG also allows  for concurrency of task execution in the backward phases. Once the gradient tensors g1, g2, g3 are computed, the SGD updates of the corresponding weight tensors w1, w2, w3 can progress concurrently. As discussed earlier, MXNET uses a PS and exports a Key-Value Store (KVStore) to the workers. MXNET allows for the above mentioned concurrency by assigning  different key "names" to each of the tensors and issuing "push" and "pull" operations to the PS using these names. These operations aggregate the gradients globally across all the workers. It is important to note that main thread doesn't block for any of these operations to complete. It issues these operations asynchronously to an internal engine tagging all the operations with explicit read/write dependencies derived from the DAG's topology. 


 Naively translating all the KVStore operations to MPI\_Allreduce operators leads to erroneous executions. Consider a scenario where a task from worker 0 initiates aggregation of tensor g2 while worker 1's task starts the operation on tensor g3. Using the same communicator, this results in silent deadlocks or program crashes. The same MPI ordering of collectives has to be maintained irrespective of how many multiple tasks may be spawned by a worker. In this paper, a worker is synonymous with a MPI process. Further, to allow for concurrency, multiple aggregations may proceed at the same time. This leads to an important question: How can MPI collectives be integrated into parallel DAGs ensuring correctness and at the same time allowing multiple simultaneous collective operations in progress?
 
In contemporary frameworks a topological ordering of all the nodes of the DAG is used to issue the operations. The MPI collectives can use the same ordering across all the workers. However, following topological order execution will not lead to maximum concurrency of all the tasks in the DAG. 
In this paper, we describe three different design options to embed MPI collectives in DAGs. Depending on the specific design, either one or multiple threads execute the collective operations. In scenarios employing multiple threads, the MPI semantics are followed either by employing multiple communicators or injecting MPI ordering dependency when pushing the communication tasks to the MXNET engines. This technique can avail of the dependency tracking algorithm in MXNET ensuring tasks are executed in same order across all DAGs. 

\begin{figure*}[h]
\includegraphics[width=\textwidth]{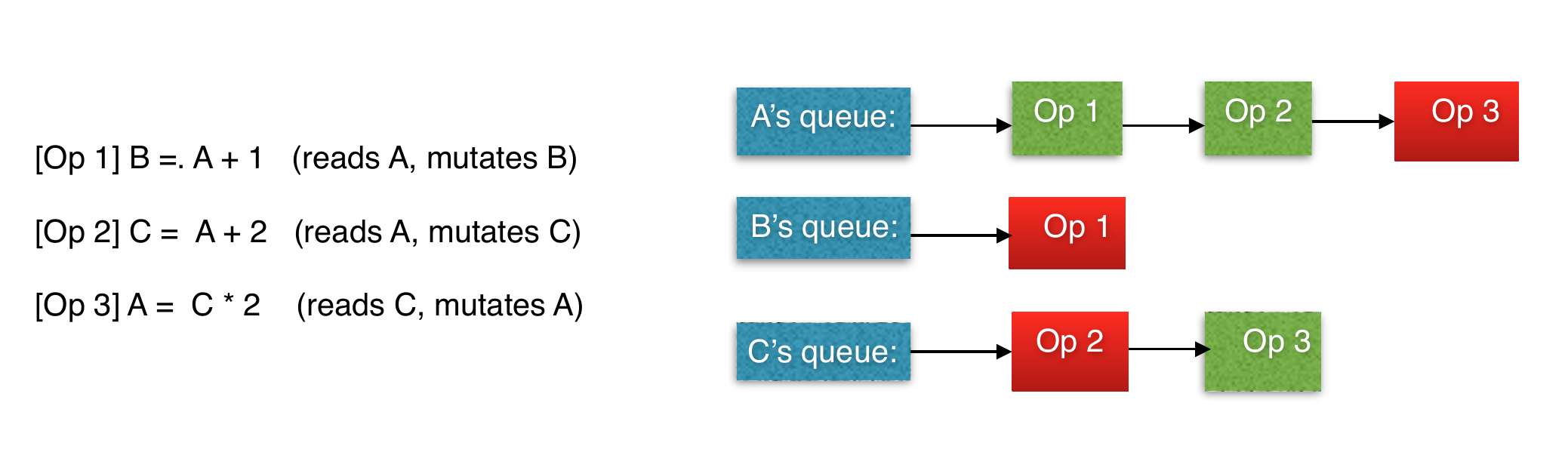}
  \caption{Dependency Engine Execution}
  \label{fig:dependency}
\end{figure*}%

\section{MXNET parallel DAG execution}\label{sec:deep-learning-using}
We first explain the dependency tracking in a single MXNET worker. We then demonstrate how this is used in conjunction with KVStore API for parallel execution of DAGs.

\subsection{Dependency Tracking }\label{sec:dependecy-tracking} In MXNET, the tasks are executed by a dependency engine based on ordering constraints. The engine is generic and it can schedule any operation ranging from GPU computations to fetching data via IO. The key to enabling this mechanism is to explicitly tag the objects in the operation that are either to be mutated via writes or to remain constant for doing reads. An operation like $b=a+1$ can be performed  by constructing the following lamda function and pushing to the engine:
\texttt{Engine.push(lambda:b.data=a.data+1, read=[a.tag], mutate=[b.tag])}.
Only the main thread issues these operation and the dependencies constructed via tagging allows the engine to generate a DAG scheduling order of these operations. This also ensures the same DAG is constructed across all parallel instances. Also, as we describe in the next section, these lamda functions become key constructs to offload MPI communications and integrate into the data flow graph.

The dependency tracking algorithm consists of separate operation queues for each object tagged.
Figure~\ref{fig:dependency} shows three separate queues for each of objects A, B and C. The basic idea in the algorithm is to enqueue the operation in all the objects' queue which make up its dependency list.
An atomic counter in the operation keeps track of the dependencies completed. Thus, if object A is ready, Ops one and two get their counters decremented by one. For read operations multiple back to back operation counters enqueued in a object's queue can be decremented all at once. Once the operation counter hits zero from all its dependency object queues, it is ready to be scheduled for execution. However, operation three has to wait until object C is written in the previous operation. C's queue tracks this dependency.
Further, successive writes get executed in order, one at a time. We use this property to impose MPI ordering semantics, explained with details in section~\ref{sec:design}.

\subsection{Task Execution}
Once the operations are scheduled, these tasks are executed on different thread pools depending on the type of the operation. For example, there is a separate thread pool to execute GPU kernels and a separate pool to copy data in and out of the GPUs. For tasks executing on CPU, such as the KVStore API, there is another dedicated pool of threads. Thus, MXNET asynchronously processes all the tasks, irrespective of their type. This allows for maximum overlap. We also use this property to break down collective operations into sub-tasks and enqueue them on the dependency queues for providing their concurrent execution, section~\ref{sec:design}.
\subsection{KVStore API }\label{sec:kvstore-api}
As described in \ref{sec:parallel-sgd}, parallel SGD needs to globally aggregate gradients across all the GPUs solving a synchronous SGD. 

\begin{figure}[h]
  \begin{lstlisting}
key= 1
gpu= 0
shape= (2,3)
g= mxnet.ndarray.ones(shape,gpu)
kvstore.init(key, g.shape)
kvstore.push(key, g)
kvstore.pull(key, g)

  \end{lstlisting}
  \caption{}
  \label{fig:code-snippet-1}
\end{figure}%
\vspace{-1mm}


The KVStore API provides distributed aggregation using PS by employing\emph{\textless{}key,
value\textgreater{}} pairs. 
Both primitives take a list of keys and list of tensors. 
The python snippet in figure \ref{fig:code-snippet-1} illustrates the basic semantics of the operation. As shown in the figure, a tensor with the gradient's shape is initialized in line 5 on the server with an alloted key and then push, pull operations are performed on it using the same key. In the DL neural net, MXNET linearly orders all the relevant tensors and assigns unique keys, starting from zero. 
Importantly, the push operation (line 6) aggregates all the gradients of the same key on the distributed KVStore servers.

The execution of these operations is handled similarly to other tasks, section~\ref{sec:dependecy-tracking}. The push and pull API are "offloaded" to the dependency engine with all the relevant dependencies. 
For example, in push (line 6), the object "g" is added to the read dependency list of push but added to the write dependency of the pull (line 7). 
The final operation would be updating the model weights using the gradient obtained from pull. 

\section{Embedding MPI operators in DAGs}\label{sec:design}
As discussed earlier in section~\ref{sec:motivation}, the two objectives of the designs discussed in the paper are 1) correctly embed MPI collective operations in the DAG and 2) overlap these operations with rest of the tasks as the DAGs offer multiple sources of concurrency, ranging from the underlying architecture of the neural nets to the different types of tasks getting executed. We now explain how the designs described in this section achieve them. We describe three different designs, introduced in Section~\ref{sec:introduction}: Funneled (Funnel), Dependency Chaining (DepCha) and Concurrent Communicators (ConCom).
All of the designs consist of two components:1) Higher level model parallelism expressed as python code and 2) Low level system implementation of the primitives in C++. 

Moreover, we also need to transition from parameter server to MPI only semantics. We accomplish this by adding a "MPI" type to the already existing types of kvstore and passing it to \texttt{KVStore.create("type")}. This type can be either, "funnel", "depCha" or "conCum". In all the implementations, the workers call MPI\_Init() or MPI\_Init\_thread() to create  MPI\_COMM\_WORLD. Further the rank of the worker in original PS is the same as MPI\_Comm\_rank() in MPI\_COMM\_WORLD. Also, in this paper, we assume each worker owns a single GPU though MXNET supports more than one GPU per worker. Since there are no servers, \texttt{KVStore.init(key, shape)} translates into the C++11 code snippet in Figure. The lambda function "offloads" MPI communication to the dependency engine in a manner similar to as shown in section ~\ref{sec:deep-learning-using}. As shown in the function, MPI process with rank zero uses MPI\_Bcast to initialize the weights across all the workers. For the sake of brevity, we do not strictly adhere to the exact MXNET APIs. Our intent is to capture the semantics of the operation rather than the syntax.

\begin{figure}[H]
  \begin{lstlisting}
auto initialize_key = [this, key, vals]() {
   MPI_Bcast(vals.data, vals.len, vals.type,0, comm);
}
Engine.Push(initialize_key, read_deps(vals.tag), mutate(none));
  \end{lstlisting}
  \caption{}
  \label{fig:kvstore-init}
\end{figure}%

\subsection{MPI Funneled}
This design solves the ordering problem of MPI collectives by making the main thread do all the communication. Figure~\ref{fig:funnel-model} shows the high level python code that summarizes the model execution. After the executor finished "offloading" the forward and backward functions to the dependency engine in line 4, lines 6-10 deal with the parallelization. In particular, the keys are ordered and push, pull operations are applied on them to aggregate the gradients indexed by key. An update is issued next. This corresponds to the operations shown in figure~\ref{fig:parallel-dag}. The global operator, MPI\_Allreduce is used in push instead of the original PS operation, Figure~\ref{fig:funnel-push}. However, though the operations issued are in order, there is no guarantee that MPI\_Allreduce will be executed in the same order on two separate workers. This is easily possible as two different collectives can be picked up by two separate threads for final execution. 
\begin{figure}[H]
  \begin{lstlisting}
Kvstore.Create("funnel")
for epoch in range(num_epochs):
 for batch in train_data:
    Executor. Forward_backward(net.symbol, net.params,
                             net.grads, batch) 
    for key in range(num_tensors):
      Kvstore.Push(key, net.grads[key])
      Kvstore.Pull(key, net.grads[key])
      SGD.Update(net.params[key], net.grads[key],
                   rescale=1/mini_batch_size)
Executor.ValidationAccuracy(test_batch)
  \end{lstlisting}
  \caption{Funneled (Python)}
  \label{fig:funnel-model}
\end{figure}%
Thus, we enforce ordering explicitly by making the main thread wait till a gradient is computed. We do this in the push (line 3, Figure~\ref{fig:funnel-push}). Also, since the main thread does MPI\_Allreduce after the wait, the MPI ordering semantics are followed. 
Making the main thread block until gradient finishes computation doesn't impede the overlap as other threads continue to pull tasks from the scheduling queue and make independent progress. Also, note that we do all the MPI operations from communication buffers( comm\_buf). 

\begin{figure}[H]
  \begin{lstlisting}
auto send_buf = comm_buf[key];
CopyFromTo(g, send_buf);
send_buf.WaitToRead();
MPI_Allreduce(MPI_IN_PLACE, send_buf.data(), send_buf.size(),
 		MPI_FLOAT, MPI_SUM, MPI_COMM_WORLD);
  \end{lstlisting}
  \caption{Funneled: KVStore Push (C++)}
  \label{fig:funnel-push}
\end{figure}%
Figure~\ref{fig:funnel-pull}, shows the copy routine to copy the gradient accumulated in the comm\_buf to the gradient tensor, g on the GPU. The copy routine internally adds the required dependencies so that the engine can schedule the SGD update step: line 9, Figure~\ref{fig:funnel-model} after the copy operation is done.
\begin{figure}[H]
  \begin{lstlisting}
auto recv_buf = comm_buf[key];
CopyFromTo(recv_buf, g);
  \end{lstlisting}
  \caption{Funneled: KVStore pull(C++)}
  \label{fig:funnel-pull}
\end{figure}%

\subsection{MPI Concurrent Communications}
In the earlier design, only one MPI collective communication can be outstanding at one time, since only the main thread invokes communication. In this design, more than one MPI\_Allreduce can make progress at a time. We accomplish this by creating multiple communicators and hashing the keys to these communicators. This strategy allows the main thread to delegate collective communication to helper threads. An important design parameter is the number of communicators required. Ideally, the number of communicators will be the same as the total number of keys or gradient tensors to be aggregated. Practically, however, the determining factor is the number of worker threads spawned for communication. Thus, we can design the python module as shown in the Figure~\ref{fig:conCum-model}.
\begin{figure}[H]
  \begin{lstlisting}
Kvstore.Create("ConCum")
OUTSTANDING=2
for epoch in range(num_epochs):
  for batch in train_data:
    Executor.Forward_backward(net.symbol, net.params,
                                  net.grads, batch)
    for key in range(num_tensors):
      Kvstore.Push(key, net.grads[key])
      Kvstore.Pull(key, net.grads[key])
      SGD.Update(net.params[key], net.grads[key],
                   rescale=1/mini_batch_size)
    if (key % OUTSTANDING == 0):
      Kvstore.barrier()
Executor.ValidationAccuracy(test_batch)
  \end{lstlisting}
  \caption{Concurrent Collectives(Python)}
  \label{fig:conCum-model}
\end{figure}%
We use a parameter,\emph{outstanding} to limit the number of concurrent collective operations to be launched. We do this by calling a barrier after every \emph{outstanding} keys have been processed. To implement this barrier, we internally track the number of allreduce operations issued in line 8, figure~\ref{fig:conCum-push}. Also, as seen from the figure, allreduce lamda function (line 4) is pushed to the engine (line 10). Thus, it can be executed by the worker thread instead of the main thread.  The barrier call simply translates to MPI\_Barrier after waiting for \emph{mpi\_outstanding} to drop to zero. \emph{mpi\_outstanding} is incremented every time the lambda function is pushed to the engine and decremented after every MPI\_Allreduce is done (line 9).The pull function, is the same as the earlier design, figure~\ref{fig:funnel-pull}
\begin{figure}[H]
  \begin{lstlisting}
auto send_buf = comm_buf[key];
CopyFromTo(g, send_buf);
send_buf.WaitToRead();
auto allreduce = [this, key, g](){
 MPI_Allreduce(MPI_IN_PLACE, send_buf.data(),
    send_buf.size(), MPI_FLOAT, MPI_SUM, comms[key]);
 --mpi_outstanding;
}
++mpi_outstanding;
Engine.Push(allreduce,read(g.tag),mutate(send_buf.tag()));
  \end{lstlisting}
  \caption{Concurrent Collectives: KVStore Push(C++)}
  \label{fig:conCum-push}
\end{figure}%


\subsection{MPI Dependency Chaining}
All the global collectives, such as MPI\_Allreduce can be broken down into three phases: 1) reduction within a node across all GPUs, 2) allreduce of the result from a node across all the nodes and 3) broadcast of the final result across all the GPUs. 
The previous designs limited the overlap of these phases. In \emph{funneled}, until a key is pushed and pulled, the main thread doesn't pick up the next key, limiting the number of outstanding collectives to only one. This is relaxed in the following scheme, \emph{ConCum} where multiple communicators can progress. However, even in this design, the stages 1 and 2 are still coupled together and we are limited by the number of worker threads delegated for collective communication for concurrency. The third design proposed here overcomes this restriction. It decouples these stages by making changes to the high level model. As shown in Figure~\ref{fig:depCha-model}, instead of calling push and pull of a key in successive calls, we break it into two separate batch invocations containing all the keys. This way all the KVStore "pushes" of the keys are scheduled at once ahead of the "pulls". 
\begin{figure}[H]
  \begin{lstlisting}
Kvstore.Create("DepCha")
for epoch in range(num_epochs):
 for batch in train_data:
    Executor. Forward_backward(net.symbol, net.params,
                             net.grads, batch) 
    for key in range(num_tensors):
      Kvstore.Push(key, net.grads[key])
    for key in range(num_tensors):
      Kvstore.Pull(key, net.grads[key])
      SGD.Update(net.params[key], net.grads[key],
                   rescale=1/mini_batch_size)
Executor.ValidationAccuracy(test_batch)
  \end{lstlisting}
  \caption{Dependency Chaining: Python}
  \label{fig:depCha-model}
\end{figure}%
The low level implementations of the push copies data into the comm buffer from the GPU, figure~\ref{fig:depCha-push}. This is suffice since we pair only one GPU to a MPI process. If there is more than one GPU attached to a process, libraries like NCCL can be used to do the reductions across GPUs in this operation. The next two stages, network reduction and local broadcast translate into MPI\_Allreduce and copying back the aggregated gradient from the comm buffer to GPU. These two stages are coded in the pull, figure~\ref{fig:depCha-pull}. 
\begin{figure}[H]
  \begin{lstlisting}
auto send_buf = comm_buf[key];
CopyFromTo(g, send_buf);
  \end{lstlisting}
  \caption{Dependency Chaining: KVStore Push(C++)}
  \label{fig:depCha-push}
\end{figure}%

A critical design parameter to consider is MPI ordering while pushing the allreduce lambda function, line 7, figure~\ref{fig:depCha-pull}. Since the engine can schedule it in different order across the workers, we ensure the same order of execution by using a \emph{dummy} tensor. We add this to the write or mutate dependency list of the operation. In MXNET, writes with respect to a tensor are always scheduled in order. 
\begin{figure}[H]
  \begin{lstlisting}
auto recv_buf = comm_buf[key];
auto allreduce = [this, key, g](){
MPI_Allreduce(MPI_IN_PLACE, send_buf.data(),
    send_buf.size(), MPI_FLOAT, MPI_SUM, 
    MPI_COMM_WORLD);
CopyFromTo(recv_buf, g);
}
Engine.Push(allreduce,read(recv_buf.tag),
            mutate(g.tag(), dummy.tag()));
  \end{lstlisting}
  \caption{Dependency Chaining: KVStore Pull(C++)}
  \label{fig:depCha-pull}
\end{figure}%

In ~\cite{s-caffe}, the authors also use threaded progress to overlap computation and communication. Their approach is specific to the implementation and doesn't use dependency tracking.

\section{EVALUATION}\label{sec:evaluation}

In this section, we explain the training efficiencies obtained using the new model of MPI + PS on CIFAR and ImageNet 1K. 
CIFAR data set comprises of 50000 training images and 10000 test images. ImageNet 1K contains about 1.2M images with 1000 classes. The total size of the training data used is 336GB and test data is 13GB. We use the latest network, Residual networks(resnet)~\cite{ResNet} using 50 layers (Resnet-50) for image classification for CIFAR data sets. For ImageNet 1K, we use Resnet-50 and Inception-BN. We compare the three designs proposed in the paper and run the models on increasing number of GPUs. 
In the figures, the legends used are:\\
\begin{itemize}[leftmargin=*]
\item Funnel: This is the MPI Funneled design proposed in section~\ref{sec:design}. In this design, the main thread ensures ordering of MPI\_Allreduce operations by blocking till the gradients are ready for global aggregation.
\item DepCha: This refers to the MPI dependency chaining proposed in section~\ref{sec:design}. In this design, all the MPI\_Allreduce operators are "offloaded" to the worker threads. The ordering is guaranteed by inserting write dependencies using a "dummy" variable created only for this purpose.
\item ConCum: In this design, multiple MPI communicators are used and keys are hashed to them to allow concurrency in MPI\_Allreduce operations. More than one thread runs these in parallel. 
\end{itemize}

\textbf{testbed}: Comprises of 64 IBM Minsky Power8 nodes with 4 NVIDIA Pascal GPUs on each node connected with InfiniBand CX5 adapters. On this testbed, we demonstrate the tensor collectives and also show the scaling behavior of ImageNet training using the optimizations proposed.



The following metrics are used to measure the performance of our
approaches:

\begin{enumerate}
\def\labelenumi{\alph{enumi})}
\item
  Epoch Time: It's time taken by the workers to train the model over the
  mini-batches of the epoch assigned to it. For multiple workers, we
  take the average time over all the workers.
\item
  Validation Accuracy: The accuracy obtained by using the model on the the separate test samples, done after every epoch. 
\end{enumerate}

\subsection{CIFAR dataset}
\begin{figure}[h]
\centering
\includegraphics[width=\columnwidth]{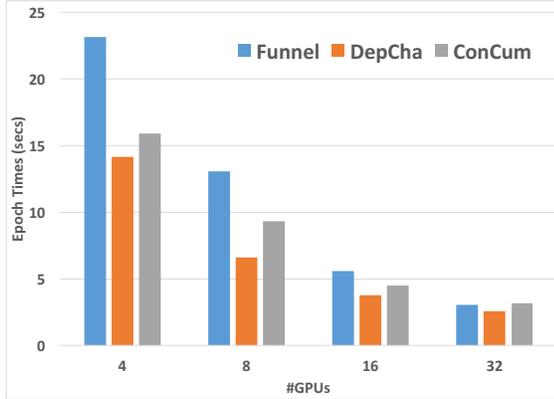}
  \caption{CIFAR Avg Epoch time, Resnet-50 (seconds)}
  \label{fig:cifar-avg-epoch-time-resnet}
\end{figure}

Figure~\ref{fig:cifar-avg-epoch-time-resnet} shows the average epoch time of training the datasets on increasing number of GPUs. As shown in the figure, across all the configurations, the dependency chaining design outperforms all the rest. However, the performance gap is widest at 8 GPUs and at 32 GPUs, we don't see significant improvements in using the dependency chaining. CIFAR training is not very computation intensive. As we increase the number of GPUs, communication quickly becomes the bottleneck. The designs proposed in the paper, essentially aim to overlap the communication with computation by allowing for concurrency across the different tasks of the DAG. At GPU counts of 32, there is not enough compute time that can be overlapped with communication time in MPI\_Allreduce. Hence, all the designs perform equally well. In contrast, ImageNet discussed next is computationally expensive providing ample room for hiding communication. 

\subsection{ImageNet 1K dataset}

\begin{figure}[h]
\centering
\includegraphics[width=\columnwidth]{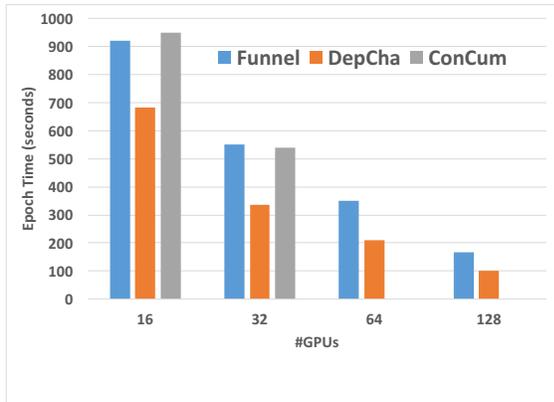}
  \caption{Imagenet Avg Epoch time, Inception-BN (seconds)}
  \label{fig:imagenet-avg-epoch-time-inception}
\end{figure}%

\begin{figure}[h]
\centering
\includegraphics[width=\columnwidth]{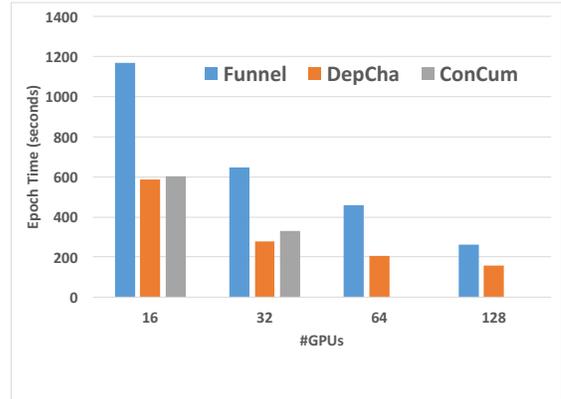}
  \caption{Imagenet Avg Epoch time, Resnet-50 (seconds)}
  \label{fig:imagenet-avg-epoch-time-resnet}
\end{figure}%

\begin{figure}[h]
\centering
\includegraphics[width=\columnwidth]{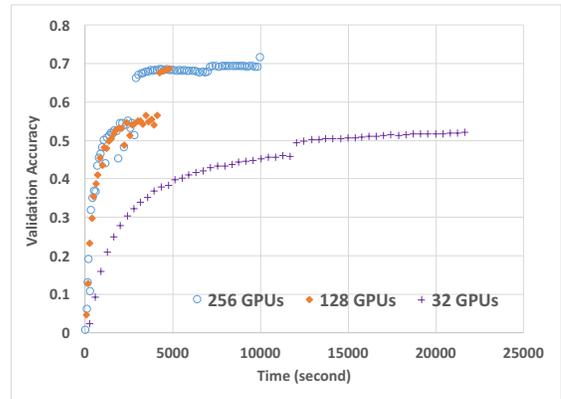}
  \caption{Imagenet Training, Resnet-50}
  \label{fig:imagenet-training-resnet}
\end{figure}%

Figure~\ref{fig:imagenet-avg-epoch-time-inception} shows the average epoch times for the Inception-BN model with the different designs proposed in the paper. We see that dependency chaining lowers the training time by a factor of more than 1.6 compared to the funneled approach for all the configurations up to 128 GPUs. Using multiple communicators for concurrent MPI\_Allreduce's across different keys also performs better than funneled. However, dependency chaining performs the best of all as it divides the collective into sub tasks and asynchronously progress each one of them. We see similar trends using Resnet with 50 layers, Figure~\ref{fig:imagenet-avg-epoch-time-resnet}.

We show the scalability of our designs in Figure~\ref{fig:imagenet-training-resnet}. Our model scales up to more than 256 GPUs with training times as low as 50 seconds per epoch. We approach over 0.71 validation accuracies with the model trained on 256 GPUs. Since the mini-batch size linearly increases with the batch size, we correspondingly increase the initial learning rate which is 0.1 by default. For 256 GPUs we set it to 1.0. 
\section{CONCLUSION}\label{conclusion}
In this paper, we described designs that allow to correctly incorporate MPI operators such as MPI\_Allreduce in parallel DAGs. Though MXNET is used to describe the various implementations, the designs are generic and can be incorporated in any DL DAG framework. 
Also, by embedding MPI into the python modules, the framework allows the user to focus on the algorithms and not deal with explicit MPI parallelization. In this aspect, it is similar to the existing popular frameworks such as Spark~\cite{kim2016deepspark, spark}, Hadoop~\cite{hadoop}.
Moreover, using MPI as the communication glue offers portability and performance for distributed DL optimizations across system and hardware architectures.



\bibliographystyle{ACM-Reference-Format}
\bibliography{paper}                     

\end{document}